\documentclass[Physsubmission, Phys]{SciPost}

\hypersetup{
    colorlinks,
    linkcolor={red!50!black},
    citecolor={blue!50!black},
    urlcolor={blue!80!black}
}

\usepackage[bitstream-charter]{mathdesign}
\DeclareSymbolFont{usualmathcal}{OMS}{cmsy}{m}{n}
\DeclareSymbolFontAlphabet{\mathcal}{usualmathcal}

\def\beq{\begin{equation}}
\def\eeq{\end{equation}}
\def\beqa{\begin{eqnarray}}
\def\eeqa{\end{eqnarray}}

\begin{document}

\begin{center}{\Large \textbf{
Three-loop soft anomalous dimensions in QCD\\
}}\end{center}

\begin{center}
Nikolaos Kidonakis\textsuperscript{$\star$}
\end{center}

\begin{center}
Kennesaw State University, Kennesaw, GA 30144, USA
\\
* nkidonak@kennesaw.edu
\end{center}

\begin{center}
\today
\end{center}

\definecolor{palegray}{gray}{0.95}
\begin{center}
\colorbox{palegray}{
  \begin{tabular}{rr}
  \begin{minipage}{0.1\textwidth}
    \includegraphics[width=35mm]{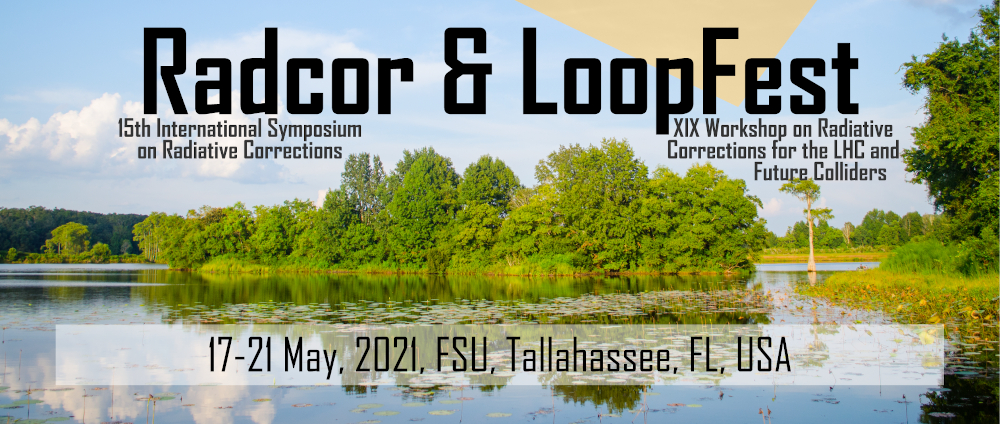}
  \end{minipage}
  &
  \begin{minipage}{0.85\textwidth}
    \begin{center}
    {\it 15th International Symposium on Radiative Corrections: \\Applications of Quantum Field Theory to Phenomenology,}\\
    {\it FSU, Tallahasse, FL, USA, 17-21 May 2021} \\
    \doi{10.21468/SciPostPhysProc.?}\\
    \end{center}
  \end{minipage}
\end{tabular}
}
\end{center}

\section*{Abstract}
{\bf
I present results for soft anomalous dimensions through three loops for many QCD processes. In particular, I give detailed expressions for soft anomalous dimensions in various processes with electroweak and Higgs bosons as well as single top quarks and top-antitop pairs.
}

\vspace{10pt}
\noindent\rule{\textwidth}{1pt}
\tableofcontents\thispagestyle{fancy}
\noindent\rule{\textwidth}{1pt}
\vspace{10pt}

\section{Introduction}

The calculation of higher-order soft-gluon corrections in perturbative QCD requires calculations of soft anomalous dimensions, $\Gamma_S$, for the corresponding processes \cite{NKGS}. The current state-of-the-art for $\Gamma_S$ for many processes is three loops. In this paper, I present results for $\Gamma_S$ for various processes at hadron colliders. These include processes with $W$, $Z$, $\gamma$, and $H$ bosons, as well as single-top and top-pair production, and $2 \to 3$ processes involving top quarks produced in association with electroweak or Higgs bosons.

Soft-gluon corrections are very important because they are typically large and they dominate the perturbative corrections for a multitude of processes, especially those involving top quarks. We consider partonic processes $p_a +p_b \to p_1 +p_2+ \cdots$
and define $s=(p_a+p_b)^2$, $t=(p_a-p_1)^2$, $u=(p_b-p_1)^2$ and $s_4=s+t+u-\sum m_i^2$. At partonic threshold $s_4 \rightarrow 0$,
and the soft corrections at order $\alpha_s^n$ involve logarithmic terms of the form $\ln^k(s_4/M^2)/s_4$, with $M$ a hard scale and $k \le 2n-1$. In order to resum these soft corrections in the (differential) cross section at NLL, NNLL, and N$^3$LL accuracy, we need 
to calculate soft anomalous dimensions at, correspondingly, one loop, two loops, and three loops.

If we take transforms of the cross section, with transform variable $N$, then we can write a factorized expression as
\beq
\sigma^{ab \to 12\cdots}(N)
= {\rm tr} \left\{H^{ab \to 12\cdots} \, 
S^{ab \to 12\cdots}\left(\frac{\sqrt{s}}{N \mu_F}\right) \right\} \, 
\psi_a \left(N_a,\mu_F\right) \, 
\psi_b \left(N_b,\mu_F\right) \, 
\prod J_i \left(N,\mu_F\right)
\nonumber
\eeq
where the $\psi$ and $J$ functions describe collinear emission from incoming and outgoing partons, $H^{ab \to 12\cdots}$ is a short-distance hard function, and $S^{ab \to 12\cdots}$ is a soft function which describes soft-gluon emission \cite{NKGS} and which satisfies the renormalization group equation
\beq
\left(\mu_R \frac{\partial}{\partial \mu_R}
+\beta(g_s)\frac{\partial}{\partial g_s}\right)\,S^{ab \to 12\cdots}
=-\Gamma_S^{\dagger \, ab \to 12\cdots} \, \, S^{ab \to 12\cdots}
-S^{ab \to 12\cdots} \, \, \Gamma_S^{ab \to 12\cdots} \, .
\nonumber
\eeq
The soft anomalous dimension $\Gamma_S^{ab \to 12\cdots}$ controls the evolution of the soft function which gives the exponentiation of logarithms of $N$ in the resummed cross section. For a recent review of soft anomalous dimensions for many QCD processes, see Ref. \cite{NKuni}.

\section{Cusp anomalous dimension}

The cusp anomalous dimension \cite{AMP,BNS,IKR,GKAR,NK2loop,GHKM,NK3loopcusp} is the simplest type of $\Gamma_S$ and a basic ingredient of calculations for QCD processes.
For eikonal lines with momenta $p_i$ and $p_j$ we define the cusp angle $\theta=\cosh^{-1}(p_i\cdot p_j/\sqrt{p_i^2 p_j^2})$.
The perturbative series is $\Gamma_{\rm cusp}=\sum_{n=1}^{\infty} (\alpha_s/\pi)^n \, \Gamma^{(n)}_{\rm cusp}$
where at one loop $\Gamma_{\rm cusp}^{(1)}=C_F (\theta \coth\theta-1)$, at two loops
\beqa
\Gamma_{\rm cusp}^{(2)}&=& K_2 \, \Gamma_{\rm cusp}^{(1)}
+\frac{1}{2}C_F C_A \left\{1+\zeta_2+\theta^2 
-\coth\theta\left[\zeta_2\theta+\theta^2
+\frac{\theta^3}{3}+{\rm Li}_2\left(1-e^{-2\theta}\right)\right] \right. 
\nonumber \\ && \hspace{30mm} \left.
{}+\coth^2\theta\left[-\zeta_3+\zeta_2\theta+\frac{\theta^3}{3}
+\theta \, {\rm Li}_2\left(e^{-2\theta}\right)
+{\rm Li}_3\left(e^{-2\theta}\right)\right] \right\} \, ,
\nonumber
\eeqa
and at three loops
$\Gamma_{\rm cusp}^{(3)}= K_3 \Gamma_{\rm cusp}^{(1)}
+2 \, K_2 (\Gamma_{\rm cusp}^{(2)}-K_2 \, \Gamma_{\rm cusp}^{(1)})+C^{(3)}$, 
where $K_3$ and $C^{(3)}$ have long expressions (see Refs. \cite{NKuni,NK3loopcusp} for explicit expressions) 
and $K_2=C_A (67/36-\zeta_2/2)-(5/18)n_f$.

In the case of the production of heavy-quark pairs, with mass $m$, we can also write the above expressions in terms of $\beta=\tanh(\theta/2)=\sqrt{1-(4m^2/s)}$, and denote them by 
$\Gamma_{\rm cusp}^{(n) \, \beta}$.

If eikonal line $i$ represents a massive quark and eikonal 
line $j$ a massless quark, then we have simpler expressions.
At one loop $\Gamma_{\rm cusp}^{(1) \, m_i}=C_F [\ln(2 p_i \cdot p_j/(m_i \sqrt{s}))-1/2]$, 
at two loops $\Gamma^{(2) \, m_i}_{\rm cusp}=K_2 \, \Gamma_{\rm cusp}^{(1) \, m_i}+(1/4) C_F C_A (1-\zeta_3)$, and at three loops
\beq
\Gamma^{(3) \, m_i}_{\rm cusp}=K_3 \, \Gamma_{\rm cusp}^{(1) \, m_i}
+ \frac{1}{2} K_2 C_F C_A (1-\zeta_3)
+C_F C_A^2\left(-\frac{1}{4}+\frac{3}{8}\zeta_2-\frac{\zeta_3}{8}-\frac{3}{8}\zeta_2 \zeta_3+\frac{9}{16} \zeta_5\right) \, .
\nonumber
\eeq

If both eikonal lines are massless, then
$\Gamma_{\rm cusp}^{\rm massless}=C_F \ln(2 p_i \cdot p_j/s) \, \sum_{n=1}^{\infty} (\alpha_s/\pi)^n \, K_n$.

\section{$\Gamma_S$ for some simple processes}

For processes with trivial color structure, the soft anomalous dimension is very simple. In fact 
$\Gamma_S$ vanishes for the following: 
Drell-Yan processes $q{\bar q} \to \gamma^*$, $q{\bar q} \to Z$; 
$W$-boson production via $q{\bar q'} \to W^{\pm}$; 
Higgs production via $b{\bar b} \to H$ and $gg \to H$; 
electroweak-boson pair production  $q{\bar q} \to \gamma \gamma$, $q{\bar q} \to Z Z$,  $q{\bar q} \to W^+ W^-$; 
production of two different electroweak bosons $q{\bar q} \to \gamma Z$, $q{\bar q'} \to W^{\pm} \gamma $, $q{\bar q'} \to W^{\pm} Z$; charged Higgs production via $b{\bar b} \to H^- W^+$, $b{\bar b} \to H^+ H^-$, $gg \to H^+ H^-$.

Also, for Deep Inelastic Scattering (DIS), $lq \to lq$ with subprocess $q \gamma^* \to q$, we have 
at one loop: $\Gamma_S^{(1) \, q \gamma^* \rightarrow q}=C_F \ln(-t/s)$; 
at two loops: $\Gamma_S^{(2) \, q \gamma^* \rightarrow q}=K_2 \, C_F \ln(-t/s)$;
and at three loops: $\Gamma_S^{(3) \, q \gamma^* \rightarrow q}=K_3 \, C_F \ln(-t/s)$.

More generally, when all external lines in a process are massless, then $\Gamma_S^{(2)}$ is proportional to
$\Gamma_S^{(1)}$ \cite{ADS}, but this is not true for processes with massive lines. Furthermore, at three loops for multi-leg scattering there are contributions from four-parton correlations \cite{ADG}.

\section{$\Gamma_S$ for large-$p_T$ $W$, $Z$, $\gamma$, $H$ production}

Let $V$ denote a $W$ or $Z$ boson or a photon or a Higgs boson. The soft anomalous dimension for these processes is a simple function (not a matrix) \cite{LOS,NKsv,NKRG} (see also \cite{NKuni}).

For the processes $qg \to W^{\pm}q'$, $qg \to Zq$, $qg \to \gamma q$, and $bg \to Hb$, we have 
at one loop: $\Gamma_S^{(1) \, qg \to Vq'}=C_F \ln(-u/s)
+(C_A/2) \ln(t/u)$;
at two loops: $\Gamma_S^{(2) \, qg \to Vq'}=K_2 \, \Gamma_S^{(1) \, qg \to Vq'}$; and 
at three loops: $\Gamma_S^{(3) \, qg \to Vq'}=K_3 \, \Gamma_S^{(1) \, qg \to Vq'}$.
The same $\Gamma_S$ also describes the reverse processes such as $\gamma q \rightarrow qg$.

For the processes $q {\bar q'} \to W^{\pm}g$, $q {\bar q} \to Zg$, $q{\bar q} \to \gamma g$, and $b{\bar b} \to Hg$,
we have at one loop: $\Gamma_S^{(1) \, q{\bar q'} \to Vg}=(C_A/2) \ln(tu/s^2)$;
at two loops: $\Gamma_S^{(2) \, q{\bar q'} \to Vg}=K_2 \, \Gamma_S^{(1) \, q{\bar q'} \to Vg}$;
and at three loops: $\Gamma_S^{(3) \, q{\bar q'} \to Vg}=K_3 \, \Gamma_S^{(1) \, q{\bar q'} \to Vg}$. The same $\Gamma_S$ also describes the reverse processes such as $\gamma g \rightarrow q {\bar q}$.

\section{$\Gamma_S$ for single-top production}

We continue with results for single-top production \cite{NKsingletop,NKsch,NKtWH,NKtch,NK3loop} (see also \cite{NKuni,NKdis21}.

For single-top $t$-channel production,
${\Gamma}_S^{bq \to tq'}$ is a $2 \times 2$ matrix \cite{NKsingletop,NKtch,NK3loop}. 
Using a $t$-channel singlet-octet color basis, the matrix elements are  
at one loop 
\beqa
&& \hspace{-7mm} 
{\Gamma}_{S \, 11}^{(1) bq \to tq'}=
C_F \left[\ln\left(\frac{t(t-m_t^2)}{m_t s^{3/2}}\right)-\frac{1}{2}\right] \, , \quad \quad \quad
{\Gamma}_{S \, 12}^{(1) bq \to tq'}=\frac{C_F}{2N_c} \ln\left(\frac{u(u-m_t^2)}{s(s-m_t^2)}\right) \, , 
\nonumber \\ && \hspace{-7mm}
{\Gamma}_{S \, 21}^{(1) bq \to tq'}= \ln\left(\frac{u(u-m_t^2)}{s(s-m_t^2)}\right) \, , \quad \quad
{\Gamma}_{S \, 22}^{(1) bq \to tq'}= \frac{C_A}{2}\left[\ln\left(\frac{u(u-m_t^2)}{m_t \, s^{3/2}}\right)-\frac{1}{2}\right] 
\nonumber \\ && \hspace{50mm}
{}+\left(C_F-\frac{C_A}{2}\right) \left[\ln\left(\frac{t(t-m_t^2)}{m_t \, s^{3/2}}\right)-\frac{1}{2}
+2\ln\left(\frac{u(u-m_t^2)}{s(s-m_t^2)}\right)\right] \, ,
\nonumber
\eeqa
at two loops
\beqa
&& \Gamma_{S \, 11}^{(2) bq \to tq'}= K_2 \, \Gamma_{S \, 11}^{(1) bq \to tq'}+\frac{1}{4} C_F C_A (1-\zeta_3)\, , 
\quad \quad
\Gamma_{S \, 12}^{(2) bq \to tq'}= K_2 \, \Gamma_{S \, 12}^{(1) bq \to tq'} \, ,
\nonumber \\ &&
\Gamma_{S \, 21}^{(2) bq \to tq'}= K_2 \, \Gamma_{S \, 21}^{(1) bq \to tq'} \, , \quad \quad
\Gamma_{S \, 22}^{(2) bq \to tq'}= K_2 \, \Gamma_{S \, 22}^{(1) bq \to tq'}+\frac{1}{4} C_F C_A (1-\zeta_3) \, ,
\nonumber
\eeqa
and at three loops
\beqa
&& \hspace{-5mm} \Gamma_{S \, 11}^{(3) bq \to tq'}= K_3 \, \Gamma_{S \, 11}^{(1) bq \to tq'}
+ \frac{1}{2} K_2 C_F C_A (1-\zeta_3) 
+C_F C_A^2\left(-\frac{1}{4}+\frac{3}{8}\zeta_2-\frac{\zeta_3}{8}-\frac{3}{8}\zeta_2 \zeta_3+\frac{9}{16} \zeta_5\right) \, ,
\nonumber \\ && \hspace{-5mm}
\Gamma_{S \, 12}^{(3) bq \to tq'}= K_3 \, \Gamma_{S \, 12}^{(1) bq \to tq'} + X_{12}^{(3) bq \to tq'} \, , \quad \quad 
\Gamma_{S \, 21}^{(3) bq \to tq'}= K_3 \, \Gamma_{S \, 21}^{(1) bq \to tq'} + X_{21}^{(3) bq \to tq'} \, ,
\nonumber \\ && \hspace{-5mm}
\Gamma_{S \, 22}^{(3) bq \to tq'}= K_3 \, \Gamma_{S \, 22}^{(1) bq \to tq'}
+ \frac{1}{2} K_2 C_F C_A (1-\zeta_3) 
+C_F C_A^2\left(-\frac{1}{4}+\frac{3}{8}\zeta_2-\frac{\zeta_3}{8}-\frac{3}{8}\zeta_2 \zeta_3+\frac{9}{16} \zeta_5\right) 
\nonumber \\ && \hspace{15mm}
{}+ X_{22}^{(3) bq \to tq'} \, ,
\nonumber
\eeqa
where the $X_{ij}^{(3) bq \to tq'}$
denote unknown terms from four-parton correlations in the last three matrix elements at three loops. It is important to note that due to the color structure of this process, only the first three-loop matrix element, $\Gamma_{S \, 11}^{(3) bq \to tq'}$, contributes to the N$^3$LO soft-gluon corrections; therefore, the unknown terms in the other three-loop matrix elements do not pose a problem in deriving N$^3$LO results. 

For single-top $s$-channel production,
${\Gamma}_S^{q{\bar q}' \to t{\bar b}}$ is also a $2 \times 2$ matrix \cite{NKsingletop,NKsch,NK3loop}. Using an  $s$-channel singlet-octet color basis, we have at one loop
\beqa 
&& \hspace{-8mm}
\Gamma_{S \, 11}^{(1) q{\bar q}' \to t{\bar b}}=C_F \left[\ln\left(\frac{s-m_t^2}{m_t\sqrt{s}}\right)
-\frac{1}{2}\right] \, ,
\quad \quad \quad
\Gamma_{S \, 12}^{(1) q{\bar q}' \to t{\bar b}}=\frac{C_F}{2N_c} \ln\left(\frac{t(t-m_t^2)}{u(u-m_t^2)}\right) \, , 
\nonumber \\ && \hspace{-8mm}
\Gamma_{S \, 21}^{(1) q{\bar q}' \to t{\bar b}}= \ln\left(\frac{t(t-m_t^2)}{u(u-m_t^2)}\right) \, , \quad \quad \quad
\Gamma_{S \, 22}^{(1) q{\bar q}' \to t{\bar b}}=\frac{C_A}{2} \left[\ln\left(\frac{t(t-m_t^2)}{m_t \, s^{3/2}}\right)-\frac{1}{2}\right]
\nonumber \\ && \hspace{53mm}
{}+\left(C_F-\frac{C_A}{2}\right) \left[\ln\left(\frac{s-m_t^2}{m_t \sqrt{s}}\right)-\frac{1}{2}+2\ln\left(\frac{t(t-m_t^2)}{u(u-m_t^2)}\right)\right] \, ,
\nonumber 
\eeqa
at two loops
\beqa
&&
\Gamma_{S \, 11}^{(2) q{\bar q}' \to t{\bar b}}=K_2 \, \Gamma_{S \, 11}^{(1) q{\bar q}' \to t{\bar b}}+\frac{1}{4} C_F C_A (1-\zeta_3) \, , \hspace{8mm}
\Gamma_{S \, 12}^{(2) q{\bar q}' \to t{\bar b}}=K_2 \, \Gamma_{S \, 12}^{(1) q{\bar q}' \to t{\bar b}} \, ,
\nonumber \\ &&
\Gamma_{S \, 21}^{(2) q{\bar q}' \to t{\bar b}}= K_2 \, \Gamma_{S \, 21}^{(1) q{\bar q}' \to t{\bar b}} \, , \hspace{8mm}
\Gamma_{S \, 22}^{(2) q{\bar q}' \to t{\bar b}}= K_2 \, \Gamma_{S \, 22}^{(1) q{\bar q}' \to t{\bar b}}+\frac{1}{4} C_F C_A (1-\zeta_3) \, ,
\nonumber 
\eeqa
and at three loops
\beqa 
&& \hspace{-2mm}
\Gamma_{S \, 11}^{(3) q{\bar q}' \to t{\bar b}}= K_3 \, \Gamma_{S \, 11}^{(1) q{\bar q}' \to t{\bar b}}
+\frac{1}{2} K_2 C_F C_A (1-\zeta_3)  
+C_F C_A^2\left(-\frac{1}{4}+\frac{3}{8}\zeta_2-\frac{\zeta_3}{8}-\frac{3}{8}\zeta_2 \zeta_3+\frac{9}{16} \zeta_5\right) \, ,
\nonumber \\ && \hspace{-2mm}
\Gamma_{S \, 12}^{(3) q{\bar q}' \to t{\bar b}}=K_3 \, \Gamma_{S \, 12}^{(1) q{\bar q}' \to t{\bar b}} + X_{12}^{(3) q{\bar q}' \to t{\bar b}} \, , \hspace{8mm}
\Gamma_{S \, 21}^{(3) q{\bar q}' \to t{\bar b}}= K_3 \, \Gamma_{S \, 21}^{(1) q{\bar q}' \to t{\bar b}} + X_{21}^{(3) q{\bar q}' \to t{\bar b}} \, ,
\nonumber \\ && \hspace{-2mm}
\Gamma_{S \, 22}^{(3) q{\bar q}' \to t{\bar b}}= K_3 \, \Gamma_{S \, 22}^{(1) q{\bar q}' \to t{\bar b}}
+\frac{1}{2} K_2 C_F C_A (1-\zeta_3)+C_F C_A^2\left(-\frac{1}{4}+\frac{3}{8}\zeta_2-\frac{\zeta_3}{8}-\frac{3}{8}\zeta_2 \zeta_3+\frac{9}{16} \zeta_5\right) 
\nonumber \\ && \hspace{18mm}
{}+ X_{22}^{(3) q{\bar q}' \to t{\bar b}} \, ,
\nonumber
\eeqa
where the $X_{ij}^{(3) q{\bar q}' \to t{\bar b}}$
denote unknown terms in the last three matrix elements. Again, we note that only the first three-loop matrix element, $\Gamma_{S \, 11}^{(3) q{\bar q}' \to t{\bar b}}$, contributes to the N$^3$LO soft-gluon corrections.
 
For associated $tW$ production the soft anomalous dimension is a simple function \cite{NKsingletop,NKtWH,NK3loop}.
At one loop
\beq
\Gamma_S^{(1) bg \to tW}=C_F \left[\ln\left(\frac{m_t^2-t}{m_t\sqrt{s}}\right)
-\frac{1}{2}\right] +\frac{C_A}{2} \ln\left(\frac{u-m_t^2}{t-m_t^2}\right) \, ,
\nonumber
\eeq
at two loops
\beq
\Gamma_S^{(2) bg \to tW}=K_2 \, \Gamma_S^{(1) bg \to tW}
+\frac{1}{4}C_F C_A (1-\zeta_3) \, ,
\nonumber
\eeq
and at three loops
\beq
\Gamma_S^{(3) bg \to tW}=K_3 \, \Gamma_S^{(1) bg \to tW}+\frac{1}{2} K_2 C_F C_A (1-\zeta_3)
+C_F C_A^2\left(-\frac{1}{4}+\frac{3}{8}\zeta_2-\frac{\zeta_3}{8}-\frac{3}{8}\zeta_2 \zeta_3+\frac{9}{16} \zeta_5\right) \, .
\nonumber
\eeq

The same soft anomalous dimension applies for the process $bg \to tH^-$, and for the FCNC processes, via anomalous top-quark couplings, $qg \to tZ$, $qg \to tZ'$, and $qg \to t\gamma$.

\section{$\Gamma_S$ for top-antitop pair production}

We continue with soft anomalous dimension matrices for $t{\bar t}$ production \cite{NKGS,NK2loop,FNPY,NKttbar} (see also \cite{NKuni,NKdis21}).

For top-antitop pair production via the $q{\bar q} \to t{\bar t}$ channel, 
${\Gamma}_S^{q{\bar q} \to t{\bar t}}$ is a $2 \times 2$ matrix and we use an $s$-channel singlet-octet color basis. At one loop for $q{\bar q} \to t{\bar t}$
\beqa
\Gamma^{(1) q{\bar q} \to t{\bar t}}_{S \, 11}&=&\Gamma_{\rm cusp}^{(1) \, \beta} \, ,
\quad
\Gamma^{(1) q{\bar q} \to t{\bar t}}_{12}=
\frac{C_F}{C_A} \ln\left(\frac{t-m_t^2}{u-m_t^2}\right) \, , 
\quad 
\Gamma^{(1) q{\bar q} \to t{\bar t}}_{21}=
2\ln\left(\frac{t-m_t^2}{u-m_t^2}\right) \, ,
\nonumber \\ 
\Gamma^{(1) q{\bar q} \to t{\bar t}}_{22}&=&\left(1-\frac{C_A}{2C_F}\right)
\Gamma_{\rm cusp}^{(1)} 
+4C_F \ln\left(\frac{t-m_t^2}{u-m_t^2}\right)
-\frac{C_A}{2}\left[1+\ln\left(\frac{s m_t^2 (t-m_t^2)^2}{(u-m_t^2)^4}\right)\right] \, ,
\nonumber
\eeqa
and at two loops 
\beqa
&& \hspace{-8mm} \Gamma^{(2) q{\bar q} \to t{\bar t}}_{S\, 11}=\Gamma_{\rm cusp}^{(2) \, \beta}, \, \hspace{2mm}
\Gamma^{(2) q{\bar q} \to t{\bar t}}_{12}=
\left(K_2-C_A \, N_2^{\beta}\right) \Gamma^{(1) q{\bar q} \to t{\bar t}}_{12}, \, \hspace{2mm}
\Gamma^{(2) q{\bar q} \to t{\bar t}}_{21}=
\left(K_2+C_A \, N_2^{\beta}\right) \Gamma^{(1) q{\bar q} \to t{\bar t}}_{21} \, ,
\nonumber \\ && \hspace{-8mm}
\Gamma^{(2) q{\bar q} \to t{\bar t}}_{22}=
K_2 \Gamma^{(1) q{\bar q} \to t{\bar t}}_{22}
+\left(1-\frac{C_A}{2C_F}\right)
\left(\Gamma_{\rm cusp}^{(2) \, \beta}-K_2 \Gamma_{\rm cusp}^{(1) \, \beta}\right)
+\frac{1}{4} C_A^2(1-\zeta_3) \, ,
\nonumber
\eeqa
where
\beq
N_2^{\beta}=\frac{1}{4}\ln^2\left(\frac{1-\beta}{1+\beta}\right)
+\frac{(1+\beta^2)}{8 \beta} \left[\zeta_2
-\ln^2\left(\frac{1-\beta}{1+\beta}\right)
-{\rm Li}_2\left(\frac{4\beta}{(1+\beta)^2}\right)\right] \, .
\nonumber
\eeq
At three loops for $q{\bar q} \to t{\bar t}$ we can write the last matrix element as
\beqa
\Gamma_{S \, 22}^{(3) q{\bar q} \to t{\bar t}}&=&
K_3 \, \Gamma_{S \, 22}^{(1) q{\bar q} \to t{\bar t}}
+\left(1-\frac{C_A}{2C_F}\right) \left(\Gamma_{\rm cusp}^{(3) \, \beta}-K_3 \Gamma_{\rm cusp}^{(1) \, \beta}\right)+\frac{1}{2}K_2 C_A^2(1-\zeta_3)
\nonumber \\ &&
{}+C_A^3\left(-\frac{1}{4}+\frac{3}{8}\zeta_2-\frac{\zeta_3}{8}-\frac{3}{8}\zeta_2\zeta_3+\frac{9}{16}\zeta_5\right)+X_{22}^{(3) q{\bar q} \to t{\bar t}} \, ,
\nonumber
\eeqa
where $X_{22}^{(3) q{\bar q} \to t{\bar t}}$ denotes unknown three-loop contributions from four-parton correlations.
The other matrix elements are also not fully known at three loops, but they have an analogous structure to that at two loops (essentially, replace (2)'s by (3)'s in the superscripts as well as replace $K_2$'s by $K_3$'s, and add $X$ terms for unknown contributions).

For top-antitop pair production via the $gg \to t{\bar t}$ channel, $\Gamma_S^{gg \to t{\bar t}}$ is a $3\times3$ matrix, and we use a color basis 
$c_1=\delta^{ab} \delta_{12}$, $c_2= d^{abc} T^c_{12}$, $c_3= if^{abc} T^c_{12}$.
We have
\beqa
\Gamma_S^{gg \to t{\bar t}}=\left[\begin{array}{ccc}
\Gamma_{S \, 11}^{gg \to t{\bar t}} & 0 & \Gamma_{S \, 13}^{gg \to t{\bar t}} \vspace{2mm} \\
0 & \Gamma_{S \, 22}^{gg \to t{\bar t}} & \Gamma_{S \, 23}^{gg \to t{\bar t}} \vspace{2mm} \\
\Gamma_{S \, 31}^{gg \to t{\bar t}} & \Gamma_{S \, 32}^{gg \to t{\bar t}} & \Gamma_{S \, 22}^{gg \to t{\bar t}}
\end{array}
\right] \, .
\nonumber
\eeqa
At one loop for $gg \to t{\bar t}$
\beqa
\Gamma_{S \, 11}^{(1) gg \to t{\bar t}}&=& \Gamma_{\rm cusp}^{(1) \, \beta}  \, ,
\quad
\Gamma_{S \, 13}^{(1) gg \to t{\bar t}}= \ln\left(\frac{t-m_t^2}{u-m_t^2}\right) \, , \quad 
\Gamma_{S \, 31}^{(1) gg \to t{\bar t}}= 2 \ln\left(\frac{t-m_t^2}{u-m_t^2}\right), 
\nonumber \\
\Gamma_{S \, 22}^{(1) gg \to t{\bar t}}&=&\left(1-\frac{C_A}{2C_F}\right) \Gamma_{\rm cusp}^{(1) \, \beta} 
+\frac{C_A}{2}\left[\ln\left(\frac{(t-m_t^2)(u-m_t^2)}{s\, m_t^2}\right)-1\right] \, , \quad 
\nonumber \\
\Gamma_{S \, 23}^{(1) gg \to t{\bar t}}&=&\frac{C_A}{2} \ln\left(\frac{t-m_t^2}{u-m_t^2}\right) \, , \quad
\Gamma_{S \, 32}^{(1) gg \to t{\bar t}}=\frac{(N_c^2-4)}{2N_c} \ln\left(\frac{t-m_t^2}{u-m_t^2}\right) \, ,
\nonumber
\eeqa
and at two loops
\beqa
\Gamma_{S \, 11}^{(2) gg \to t{\bar t}}&=& \Gamma_{\rm cusp}^{(2) \, \beta}, \, 
\Gamma_{S \, 13}^{(2) gg \to t{\bar t}}=\left(K_2-C_A N_2^{\beta}\right) 
\Gamma_{S \, 13}^{(1) gg \to t{\bar t}}, \,
\Gamma_{S \, 31}^{(2) gg \to t{\bar t}}=\left(K_2+C_A N_2^{\beta}\right)  
\Gamma_{S \, 31}^{(1) gg \to t{\bar t}},
\nonumber \\
\Gamma_{S \, 22}^{(2) gg \to t{\bar t}}&=& K_2 \, \Gamma_{S \, 22}^{(1) gg \to t{\bar t}}
+\left(1-\frac{C_A}{2C_F}\right) \left(\Gamma_{\rm cusp}^{(2) \, \beta}-K_2 \Gamma_{\rm cusp}^{(1) \, \beta}\right)+\frac{1}{4} C_A^2(1-\zeta_3) \, ,
\nonumber \\
\Gamma_{S \, 23}^{(2) gg \to t{\bar t}}&=& K_2 \, \Gamma_{S \, 23}^{(1) gg \to t{\bar t}} \, , \quad
\Gamma_{S \, 32}^{(2) gg \to t{\bar t}}= K_2 \, \Gamma_{S \, 32}^{(1) gg \to t{\bar t}} \, .
\nonumber
\eeqa
At three loops for $gg \rightarrow t{\bar t}$, we can write the $22$ matrix element as
\beqa
\Gamma_{S \, 22}^{(3) gg \to t{\bar t}}&=&
K_3 \, \Gamma_{S \, 22}^{(1) gg \to t{\bar t}}
+\left(1-\frac{C_A}{2C_F}\right) \left(\Gamma_{\rm cusp}^{(3) \, \beta}-K_3 \Gamma_{\rm cusp}^{(1) \, \beta}\right)+\frac{1}{2}K_2 C_A^2(1-\zeta_3)
\nonumber \\ &&
{}+C_A^3\left(-\frac{1}{4}+\frac{3}{8}\zeta_2-\frac{\zeta_3}{8}-\frac{3}{8}\zeta_2\zeta_3+\frac{9}{16}\zeta_5\right)+X_{22}^{(3) gg \to t{\bar t}} \, ,
\nonumber
\eeqa
where $X_{22}^{(3) gg \to t{\bar t}}$ denotes unknown three-loop contributions from four-parton correlations. The other matrix elements are, again, also not fully known at three loops, but they have an analogous structure to that at two loops. 

\section{$\Gamma_S$ for $tqH$, $tqZ$, $tq\gamma$, $tqW$ production}

We consider processes $bq \to tq'H$ as well as $bq \to tq'Z$, $bq \to tq'\gamma$, $bq \to tqW^-$, $qq \to tq'W^+$. We use a $t$-channel singlet-octet color basis, and we further define
$s'=(p_1+p_2)^2$, $t'=(p_b-p_2)^2$, $u'=(p_a-p_2)^2$. All these processes have the same soft anomalous dimension matrix \cite{MFNK}. We have at one loop
\beqa
&& {\Gamma}_{S \,  11}^{(1) \, bq \to tq'H}=
C_F \left[\ln\left(\frac{t'(t-m_t^2)}{m_t s^{3/2}}\right)-\frac{1}{2}\right],
\nonumber \\ &&
{\Gamma}_{S \, 12}^{(1) \, bq \to tq'H}=\frac{C_F}{2N_c} \ln\left(\frac{u'(u-m_t^2)}{s(s'-m_t^2)}\right),  
\quad
{\Gamma}_{S \, 21}^{(1) \, bq \to tq'H}= \ln\!\left(\frac{u'(u-m_t^2)}{s(s'-m_t^2)}\right) \, ,
\nonumber \\ &&
{\Gamma}_{S \,  22}^{(1) \, bq \to tq'H}= C_F \left[\ln\left(\frac{t'(t-m_t^2)}{m_t s^{3/2}}\right)-\frac{1}{2}\right]
-\frac{1}{N_c}\ln\left(\frac{u'(u-m_t^2)}{s(s'-m_t^2)}\right) 
+\frac{N_c}{2}\ln\left(\frac{u'(u-m_t^2)}{t'(t-m_t^2)}\right) \, ,
\nonumber
\eeqa
at two loops
\beqa
&& \Gamma_{S\, 11}^{(2) \,  bq \to tq'H}= K_2 \, \Gamma_{S\, 11}^{(1) \,  bq \to tq'H}
+\frac{1}{4} C_F C_A (1-\zeta_3) \, ,
\quad
\Gamma_{S\, 12}^{(2) \,  bq \to tq'H}= K_2 \, \Gamma_{S\, 12}^{(1) \,  bq \to tq'H} \, ,
\nonumber \\ &&
\Gamma_{S\, 21}^{(2) \,  bq \to tq'H}= K_2 \,  \Gamma_{S\, 21}^{(1) \,  bq \to tq'H} \, ,
\quad
\Gamma_{S\, 22}^{(2) \,  bq \to tq'H}= K_2 \,  \Gamma_{S\, 22}^{(1) \,  bq \to tq'H}
+\frac{1}{4} C_F C_A (1-\zeta_3) \, ,
\nonumber
\eeqa
and at three loops
\beqa
&& \hspace{-6mm} \Gamma_{S \, 11}^{(3) bq \to tq'H}= K_3 \, \Gamma_{S \, 11}^{(1) bq \to tq'H}
+ \frac{1}{2} K_2 C_F C_A (1-\zeta_3) 
+C_F C_A^2\left(-\frac{1}{4}+\frac{3}{8}\zeta_2-\frac{\zeta_3}{8}-\frac{3}{8}\zeta_2 \zeta_3+\frac{9}{16} \zeta_5\right) ,
\nonumber \\ && \hspace{-6mm}
\Gamma_{S \, 12}^{(3) bq \to tq'H}= K_3 \, \Gamma_{S \, 12}^{(1) bq \to tq'H} + X_{12}^{(3) bq \to tq'H} \, , \quad \quad 
\Gamma_{S \, 21}^{(3) bq \to tq'H}= K_3 \, \Gamma_{S \, 21}^{(1) bq \to tq'H} + X_{21}^{(3) bq \to tq'H} \, ,
\nonumber \\ && \hspace{-6mm}
\Gamma_{S \, 22}^{(3) bq \to tq'H}= K_3 \, \Gamma_{S \, 22}^{(1) bq \to tq'H}
+ \frac{1}{2} K_2 C_F C_A (1-\zeta_3) 
+C_F C_A^2\left(-\frac{1}{4}+\frac{3}{8}\zeta_2-\frac{\zeta_3}{8}-\frac{3}{8}\zeta_2 \zeta_3+\frac{9}{16} \zeta_5\right) 
\nonumber \\ && \hspace{15mm}
{}+ X_{22}^{(3) bq \to tq'H} \, ,
\nonumber
\eeqa
where the $X_{ij}^{(3) bq \to tq'H}$ denote unknown terms in the last three matrix elements which, however, do not contribute to the soft-gluon corrections at N$^3$LO.

We next consider the processes $q{\bar q'} \to t {\bar b} H$ as well as $q{\bar q'} \to t{\bar b}Z$, $q{\bar q'} \to t{\bar b}\gamma$, $q{\bar q} \to t{\bar b}W^-$, $q{\bar q'} \to t{\bar q''}W^+$, which all have the same soft anomalous dimension matrix \cite{MFNK}, and we use an $s$-channel singlet-octet color basis. We have at one loop
\beqa
&& \Gamma_{S \, 11}^{(1) \, q{\bar q'} \to t{\bar b}H}=C_F \left[\ln\!\left(\frac{s'-m_t^2}{m_t\sqrt{s}}\right)-\frac{1}{2}\right],
\nonumber \\ &&
\Gamma_{S \, 12}^{(1) \, q{\bar q'} \to t{\bar b}H}=\frac{C_F}{2N_c} \ln\left(\frac{t'(t-m_t^2)}{u'(u-m_t^2)}\right),  \quad \quad
\Gamma_{S \, 21}^{(1) \, q{\bar q'} \to t{\bar b}H }= \ln\!\left(\frac{t'(t-m_t^2)}{u'(u-m_t^2)}\right) \, ,
\nonumber \\ &&
\Gamma_{S \, 22}^{(1) \, q{\bar q'} \to t{\bar b}H}=C_F \left[\ln\left(\frac{s'-m_t^2}{m_t \sqrt{s}}\right)-\frac{1}{2}\right]
-\frac{1}{N_c}\ln\left(\frac{t'(t-m_t^2)}{u'(u-m_t^2)}\right)
+\frac{N_c}{2} \ln\left(\frac{t'(t-m_t^2)}{s(s'-m_t^2)}\right) \, ,
\nonumber
\eeqa
at two loops
\beqa
&& \Gamma_{S\, 11}^{(2) \,  q{\bar q'} \to t{\bar b}H}= K_2 \, \Gamma_{S\, 11}^{(1) \,  q{\bar q'} \to t{\bar b}H}+\frac{1}{4} C_F C_A (1-\zeta_3) \, ,
\quad
\Gamma_{S\, 12}^{(2) \,  q{\bar q'} \to t{\bar b}H}= K_2 \, \Gamma_{S\, 12}^{(1) \,  q{\bar q'} \to t{\bar b}H} \, ,
\nonumber \\ &&
\Gamma_{S\, 21}^{(2) \,  q{\bar q'} \to t{\bar b}H}= K_2 \,  \Gamma_{S\, 21}^{(1) \,  q{\bar q'} \to t{\bar b}H} \, ,
\quad
\Gamma_{S\, 22}^{(2) \,  q{\bar q'} \to t{\bar b}H}= K_2 \,  \Gamma_{S\, 22}^{(1) \,  q{\bar q'} \to t{\bar b}H}+\frac{1}{4} C_F C_A (1-\zeta_3) \, ,
\nonumber
\eeqa
and at three loops
\beqa
&& \hspace{-6mm} \Gamma_{S \, 11}^{(3) q{\bar q}' \to t{\bar b}H}= K_3 \, \Gamma_{S \, 11}^{(1) q{\bar q}' \to t{\bar b}H}
+\frac{1}{2} K_2 C_F C_A (1-\zeta_3)  
+C_F C_A^2\left(-\frac{1}{4}+\frac{3}{8}\zeta_2-\frac{\zeta_3}{8}-\frac{3}{8}\zeta_2 \zeta_3+\frac{9}{16} \zeta_5\right) ,
\nonumber \\ && \hspace{-6mm}
\Gamma_{S \, 12}^{(3) q{\bar q}' \to t{\bar b}H}=K_3 \, \Gamma_{S \, 12}^{(1) q{\bar q}' \to t{\bar b}H} + X_{12}^{(3) q{\bar q}' \to t{\bar b}H} \, , \hspace{8mm}
\Gamma_{S \, 21}^{(3) q{\bar q}' \to t{\bar b}H}= K_3 \, \Gamma_{S \, 21}^{(1) q{\bar q}' \to t{\bar b}H} + X_{21}^{(3) q{\bar q}' \to t{\bar b}H} \, ,
\nonumber \\ && \hspace{-6mm}
\Gamma_{S \, 22}^{(3) q{\bar q}' \to t{\bar b}H}= K_3 \, \Gamma_{S \, 22}^{(1) q{\bar q}' \to t{\bar b}H}
+\frac{1}{2} K_2 C_F C_A (1-\zeta_3)+C_F C_A^2\left(-\frac{1}{4}+\frac{3}{8}\zeta_2-\frac{\zeta_3}{8}-\frac{3}{8}\zeta_2 \zeta_3+\frac{9}{16} \zeta_5\right) 
\nonumber \\ && \hspace{15mm}
{}+ X_{22}^{(3) q{\bar q}' \to t{\bar b}H} \, ,
\nonumber 
\eeqa
where the $X_{ij}^{(3) q{\bar q}' \to t{\bar b}H}$ denote unknown terms in the last three matrix elements which, however, do not contribute to the soft-gluon corrections at N$^3$LO.

\section{Conclusion}

Soft anomalous dimensions are fundamental in describing soft-gluon emission in QCD processes. In this contribution, I presented results for soft anomalous dimensions for many processes through three loops. 
These results are needed in calculations of high-order corrections.

\paragraph{Funding information}
This material is based upon work supported by the National Science Foundation under Grant No. PHY 2112025.

\end{document}